\documentstyle[pra,aps,epsf,multicol]{revtex}

\begin{document}
\draft
\title{What does an observed quantum system reveal to its observer ?}
\author{Peter E. Toschek and Christof Wunderlich}
\address{Institut f\"ur Laser-Physik, Universit\"at Hamburg,
Jungiusstr. 9, D-20355 Hamburg, Germany}
\date{March 2, 2001}
\maketitle

\begin{abstract}
The evolution of a quantum system under observation becomes retarded or even
impeded. We review this ``quantum Zeno effect'' in the light of the
criticism that has been raised upon a previous attempt to demonstrate it, of
later reexaminations of both the projection postulate and the significance
of the observations, and of the results of a recent experiment on an {\it %
individual} cold atom. Here, the micro-state of the quantum system gets
unveiled with the observation, and the effect of measurement is no longer
mixed up with dephasing the object's wave function by the reactive effect of
the detection. A procedure is outlined that promises to provide, by
observation, an upper limit for the delay of even an {\it exponential }decay.
\end{abstract}


\begin{multicols}{2}

\section{Introduction}
An isolated quantum system found in one of its eigenstates of energy starts
evolving as soon as it is perturbed: This system when supplemented by the
perturbing entity no longer resides in one of its eigenstates, and it
becomes time-dependent. This phenomenon seems close to the behaviour of a
macroscopic system, say of a pair of coupled pendula, with its split normal
modes. However, an attempt of measuring an observable of the quantum system
brings to the fore a fundamental difference: Here, one of the eigenvalues is
the result of the measurement. Although we have become accustomed with this
fundamental teaching of quantum mechanics, still alien seems the unavoidable
conclusion that the system has suddenly turned into the eigenstate
corresponding to the observed eigenvalue. This ''state reduction'' according
to the projection postulate of von Neumann [1] and L\"{u}ders [2] is thought
to accompany the observation of the system and constitutes, as it seems, a
discontinuous break of the otherwise continuous quantum evolution. Repeated
measurement, with same result, on a quantum system amounts to reiterated
resettings of the evolving wave function. It brings about retardation or
even impediment of the evolution [3,4], when the temporal separation of the
detections is reduced, and the interaction more and more approaches
continuous measurement. This theoretical consequence, i.e. quantum evolution
impeded by measurement, or the ``quantum Zeno effect'' [QZE, 5], is based on
the likeliness of finding the system in a state {\it other} than the initial
one varying in proportion with the square of lapsed time, $(\Delta t)^2$ [3,
6-9]. With $N$ quasi-instantaneous observations of the system equally
distributed during time $T$ and separated by $T/N$, that net probability
varies as $N(\Delta t)^2=T^2/N$, whose limit at $N\rightarrow \infty $
vanishes, and with it the dynamics of the system.

An attempt of demonstrating the QZE has involved a large ensemble of atomic
particles well isolated from the environment, namely five thousand beryllium
ions located in an ion trap [10]. Another experiment that made use of light
in cascaded interferometers [11] has been shown explicable in classical
terms [8]. The ion ensemble in the trap experiment [10] interacted, aside
from the trapping field, only with two kinds of pulse of radiation: long
microwave pulses that drove the ensemble's coherent quantum evolution on a
hyperfine line, and a series of short light pulses superimposed for the
excitation of light scattering on a neighbouring resonance line, that was
assumed to represent a sequence of measurements of the system's internal
energy state.

This experiment has yielded complete agreement with the predictions of
quantum-mechanical calculations. Its analysis and the pertaining claims
have, however, aroused criticism on various counts: ({\it i}) The
interpretation of the experimental findings in terms of projection postulate
and state reduction that was said inappropriate [12, 13]. ({\it ii}) The
interpretation of these findings in terms of the quantum Zeno effect that
was said inappropriate {\it because it did not require} the application of
the projection postulate [14-16]. ({\it iii}) The recording only of the {\it %
net} probability for no transition, after a {\it series} of short light
pulses applied to the ions, in spite of the application of every individual
pulse being considered a ``measurement'' [17]. ({\it iv}) The results of the
observations being considered not to demonstrate a non-local,
negative-result effect [8]. ({\it v}) The use, for the quantum system, of a
large ensemble that makes one identify the effect of the measurements with
physically dephasing the wave function [18-23]. And ({\it vi}) the
demonstration of the perturbed evolution on a {\it coherent} dynamics, as
opposed to spontaneous, exponential decay [24].

The review of Home and Whitaker [8] gives a very detailed account of many
relevant problems and reevaluates the interpretations. It is the purpose of
the present paper to review the positions characterized by the arguments
listed above in the light of this and other reevaluations, and of a recent
experiment on an{\it \ individual} atomic system [25]. We feel that this
contribution to the debate is well justified since, after all, the QZE seems
to characterize a non-local correlation of the quantum system with a
macroscopic meter, as Bell's inequalities characterize the non-local
correlation of two quantum systems [8].

\section{Projection postulate and state reduction}

A principal line of criticism of the experiment of Itano {\it et al.} [10]
is concerned with the application of projection postulate and state
reduction to the interpretation [12, 13, 19]: A particular degree of freedom 
$-$ the weak resonance of an atomic two-level system being excited by
microwave used as the drive $-$ was said to have been inappropriately
singled out. When the probe transition -- the resonance line -- were
included in the model, as, e.g., in a set of {\it three-}level Bloch
equations, the mean evolution of the entire system is said derivable
complete with situations of almost vanishing non-diagonal elements of the
density matrix. There is no need to separately invoke projection and state
reduction, which in fact are inferred not to happen [13].

The reasoning suggests that in fact the concept of state reduction is not
required for the formal description of the dynamics of the driven quantum
system extended by what has been named ''quantum probe'' [21], and of the
relationship of driven resonance and probe resonance. However, the
photo-electric detection of the scattered light still reduces the extended
system into the ``on'' state, and failing to detect scattering reduces it
into the ``off'' state. Reduction of the extended system is now achieved by
the next link of the causal chain so far {\it not }included in the model,
i.e. here by the photo-detection of the probe-light-excited fluorescence.
Thus, this concept seems inevitable just at the borderline separating the
modelled system (now the complete three-level system) and the rest of the
world that includes the macroscopic measuring device. Moreover, the
modelling of the multi-particle system based on the densitiy matrix results
in expectation values of the observable that correspond to ``non-selective''
measurements, i.e. observations that do not reveal the micro-state of the
system. Such a type of observation, however, is inappropriate for the
demonstration of the reactive effect of a {\it measurement} since this
effect, in such an observation, completely agrees with physically dephasing
the system's wave function [18-22]. This ambiguity of interpretation will be
discussed in Section 4. A quantum system subject to selective measurements,
i.e. observations linked with detailed information on the relevant
observables of individual quantum objects that constitute the system under
scrutiny, is suitably modelled by the quantum jump approach [26] also known
as Monte Carlo wave function calculation [27]. Here, trajectories of the
evolution of an individual quantum system, or of the sub-systems of a larger
entity, are calculated whose weighted average will eventually reproduce the
results of a corresponding calculation based on the density-matrix, e.g.,
the solutions of Bloch equations. Using the quantum jump approach, Beige and
Hegerfeldt [9] have shown that a ``good measurement'' as defined by perfect
projection and state reduction is approximated to very high precision by the
effect of a probe pulse on a resonant atom. This is so in the range of
parameters defined by $A\Omega _d/\Omega _p\ll 1$ and $\Omega _p/A\ll 1$,
where $\Omega _d,\Omega _p$ are the Rabi frequencies of the drive and probe
radiation, respectively, and $A$ is the rate of spontaneous decay on the
probe line. Related results have been derived from numerical simulations
using Bloch equations [28]. The parameters of the experiment of Itano {\it %
et al.} [10] are well inside the above regime, such that ``state reduction''
remains a very good characterization of the effect of a probe-light pulse.
Consequently, the objections on the ground of the use of this concept for
the interpretation of the experimental findings seem unfounded.

On the other hand, Home and Whitaker have argued that not state reduction
and ''collapse'' are prerequisite for the ''paradoxical'' aspects of the
quantum evolution to show up, but rather the quadratic dependence on time of
the probability for survival of the system in its initial state [8]. Thus,
objections on the ground of {\it not} using the concept of collapse seem
also irrelevant.

\section{Measurement of the net{\it \ }rate of survival
instead of the true rate}

The concept of the evolution of a quantum system to be inhibited by
reiterated measurement according to the proposal of Cook was based upon the
measurable variation of the probability that {\it no} decay has been found
throughout the interval $\delta =[$0,T$]$, where T is the overall time of
the (driven) evolution [29]. Nakazato et al. [17] have pointed out that in
the experiment on trapped Be ions [10] actually a different quantity has
been recorded, namely the probability of finding no{\it \ net} decay at time
T, while N interactions with the probe light, being considered measurements,
have happened during $\delta $. Although the results of these N
``measurements'' in principle could have been recorded, in fact they were
not; instead, the state of the system was recorded {\it past} $\delta ,$
i.e. at time T. The emergent overall results ignore processes that involve
the excitation of some ion, time-correlated with deexcitation of any other
ion. Moreover, such measurements, even when exerted upon an {\it individual}
particle, could not help unravelling the true probability of {\it no}
transition since it would include an unknown number of back-and-forth
transitions that have happened on the driven line of this single quantum
system but are left unrecorded. As a consequence, the probability for no 
{\it net} transition is at variance with the probability for no transition
whatsoever. The latter, however, is used with both the definition [5] and
the quantitative evalutation of the effect of measurement upon the evolution
of a quantum system. Although both the two probabilities approach unity with
increasing number N of probe pulses, they differ at finite N. This is,
however, the typical situation encountered with an experimental proof. In
contrast, a series of measurements, during the interval $\delta $, whose
results are being {\it individually recorded}, would lack that drawback and
permit the determination of the probability that indeed no transition (decay
or excitation) has happened during the entire time of interaction. The
recent experiment that has in fact involved the registration of {\it all}
the results of the N measurements of a sequence, exerted on a single ion
[25], will be discussed in section 5, and its possible extension to decay
processes in section 8.

\section{Basis and signature of the Quantum Zeno Effect}

Various types of observations have been suggested for the demonstration of
the QZE both in model and experiment. The early proposals refer to the decay
of an unstable quantum system [3-5]. Conventional models of such a decay
inevitably require deviations from the exponential law at long and at short
times [30, 31, 7], in particular an initial quadratic time dependence.
However, these deviations have never been observed, and the question was
raised of QZE to show up or not with a{\it \ strictly} exponential decay.
Such a decay is a classical concept. Recently, however such a quantum model
was constructed [7] on the expense of a conceptual drawback, namely the mean
energy of the (position) eigenstates not being well-defined. Other models of
the exponential decay have made use of the Wigner-Weisskopf approach that
includes the physical interaction of the quantum object with an infinite
reservoir [32]. These models feature unbounded spectra of energy eigenvalues
and lack a lowest, stable, eigenstate [4, 7]. Several recent proposals seem
based on various types of such interaction; and they demonstrate even {\it %
advanced }evolution [33, 34]. However, qualification for QZE {\it proper}
(or QZ paradox [8]) seems questionable since the paradoxical aspects of QZ
rely on the retardation of evolution in the {\it absence} of any reaction,
on the quantum object, from the environment, namely as a consequence of
non-local correlation of quantum object and meter. This situation may be
unambiguously represented with the {\it coherent} evolution of a quantum
system which consequently has become widely accepted as a model for the
demonstation of the QZE [29].

This point of view has been elaborated in the comprehensive review of Home
and Whitaker [8]. These authors stress the proximity of {\it this} QZE,
based on correlation of quantum object and macroscopic detector, and Bell's
theory, that involves the non-local correlation of two quantum objects. They
suggest, for a consequent terminology, that ``QZE'' should characterize
non-local negative result measurements on a microscopic system. Such a kind
of observation would certainly suffice for discriminating the effect of
correlation against back action and deserve the characterization as QZE: The
information provided to the observer by the measurements is {\it not }%
generated {\it by local interaction}. However, it seems to us that the
latter criterium is obeyed even by two more classes of measurements that are 
{\it not} of the negative-result type: (1) measurements free of back action
-- the so-called ''quantum non-demolition measurements'' (QND) [21] -- that
may in fact give rise to positive results, and (2) measurements whose back
action {\it demonstrably } cannot account for the surmised retarding effect,
e.g., because it is to small. From a logical point of view there is no good
reason to exclude these classes of measurements from the general type that
generates the {\it true} QZE (or, ``QZ paradox''), that might well be
considered constituting a {\it necessary} criterium.

A factual demonstration according to criterium 2 may be very cumbersome, or
even practically impossible, with a quantum system of many degrees of
freedom. As for the analysis of the experiment of Itano et al. [10], the
evaluation of Ref. 8 is still assumed to hold since that experiment fails to
meet even the more general criteria 1 and 2. The situation is quite
different with an individual quantum system when picking, for the
observable, the simplest degree of freedom, a two-level system equivalent to
a spin. Here, the back action on the quantum object to be scrutinized may
turn out restricted to the small variation of a simple wave function, i.e.,
of a modulus and/or a phase. In such a system, modification of the modulus
goes along with a variation of energy. Such a variation, in particular any
dissipation, prevents a measurement from being ''good'', but it may be
excluded by the design of the experiment. A phase perturbation is harder to
avoid. This problem will be addressed in the next two sections. A recent
experiment on a {\it single} spatially confined ion will be analysed, and we
shall show that part of the results obeys criterium 2, while another part
obeys criterium 1 {\it and} even the sufficient criterium of Ref.
[8].

\section{Dephasing of the wave function of the observed system}

Several authors have noticed that the effect of even a ``good'' measurement 
{\it on an ensemble} nonetheless completely agrees with phase diffusion of
the system's wave function by the action of the environment and/or the
measuring device (``dephasing'') [18-23]. Spiller [22] as well as Alter and
Yamamoto [23] have pointed out that the effect of measurement is
discriminated from dephasing {\it only} when the quantum object of the
measurement is an {\it individual }quantum system. This is so since here
indeed the micro-state of the measured observable becomes accessible to the
observer, in contrast with a global measurement on an ensemble. As an
example, we again consider the conceptionally simplest quantum system, a
two-level atom, isomorphic with a spin 1/2 whose symmetry is SU2 [35]. The
configuration space of such a system extends over the surface of the unit
sphere whose poles correspond to the two energy eigenstates $\pm $ $\hbar
\omega _0/2$ (Fig. 1). All the other locations on this surface represent
superposition states and exhibit a moment. Aside from the relative phase of
the two eigenstates, this state may be determined from repeated measurements
of energy upon identically preparing the quantum system, i.e., of the
projection of the state onto the z axis, $\mid 0\rangle -\mid 1\rangle $.
This is in contrast with an ensemble of such spins which requires one
measurement only and represents, in general, a mixture of states whose state
vector may terminate anywhere in the interior of the unit sphere: The
reconstruction of the micro-state from the measurement is impossible as a
consequence of incomplete knowledge, more specific: of ignoring the modulus
of the moment that indicates the temporal correlations of the two
eigenstates. Analytic [21] as well as numerical calculations [23] have
proven that the evolutive dynamics of an ensemble under reiterated
observation remains indistinguishable from an evolution subject to phase
relaxation. Consequently, an unequivocal demonstration of the quantum
evolution to be impeded or retarded by repeated {\it measurements} requires
one to pick an individual quantum object as the system to be kept under
observation.

\section{Evidence against dephasing}

Recently, an experiment on a single trapped and cooled ion has been reported
[25]: The evolution of the ion was deduced from the statistics of sequences
of {\it equal results} of measurements, each measurement consisting of
driving and probing the ion on the respective neighbouring resonances. For
this purpose, a single $^{172}Yb^{+}$ ion was spatially confined in a
miniaturized electrodynamic trap in ultra-high vacuum [36]. The ion was
laser-cooled and adjusted to the node of the radio-frequency trapping field
such that it could be considered, in good approximation, as located in
field-free space. For time intervals $\tau =2ms$, the $E2$ transition $%
S_{1/2}-D_{5/2}$ was excited by almost monochromatic blue light (1-s
bandwidth less than 500 Hz) generated from frequency-doubling the 822-nm
output of a diode laser, so that driving the ion was phase-coherent during $%
\tau $ (Fig. 2). These pulses alternated with 10-ms pulses of 369-nm probe
light, generated by frequency-doubling the light of a dye laser. This probe
light excited resonance scattering by the ion at a rate of $10^8$ photons
per second (of which some $10^4$ were detected by photon counting) appeared 
{\it only} when the ion resided in the $S_{1/2}$ ground state and was
susceptible to excitation of its dipole moment on the resonance line
considered the quantum probe. Lacking light scattering was considered the
signature of the ion being found in its {\it metastable} $D$ state [37, 38].
Trajectories of 500 measurements have been recorded, each consisting of a
drive pulse, and a probe pulse with simultaneous detection. Pairs of
measurements the first of which yields the ion in the ground state, and the
second one in the metastable state, signal an act of excitation by the drive
light. From the number of such pairs observed in a trajectory, the
probability of excitation was derived. Trajectories recorded at the driving
light being stepwise scanned across the pertaining line allow one to
generate an absorption spectrum of the ion's $E2$ resonance. A corresponding
spectrum calculated from a numerical solution of Bloch's equations is shown
in Fig. 3. The probability of excitation oscillates according to the angle
of optical nutation $\theta (t)=\sqrt{\Omega ^2+\Delta ^2}$ $t$ that depends
on the Rabi frequency $\Omega $ and the detuning of the driving light, $%
\Delta =\omega -\omega _0,$ where $\omega ,\omega _0$ are the light and $E2$
resonance frequencies, respectively. The envelope of this absorption line is
characterized by the Rabi nutation frequency, whereas pulse length of the
drive and the additional broadening by the rate of dephasing, $\gamma ,$
determine the contrast of the modulation. Marked are those data points in
the spectrum whose trajectories of measurement have been subjected to
statistical evaluation. The recorded version of the excitation spectrum [25]
agrees with Fig. 3 and shows indeed the probability of excitation being Rabi
modulated. This agreement reveals the interaction of driving light and ion
to have been {\it coherent} in the experiment.

For the statistical evaluation of the results, the ion was assumed to occupy
its ground state, when probe light was about to be scattered. Then, another
such ``on'' event takes place with probability $p_0=\cos ^2(\Omega \tau /2)$%
, provided that the drive light is tuned to resonance $(\Delta =0)$. With
the ion in its metastable state, the probability for another ``off'' result
is the same, $p_1=p_0=p$, as long as relaxation is neglected. Finding a
series of $q$ equal results right after each other has the conditional
probability 
\begin{equation}
U(q)=U(1)V(q-1)\text{ ,}  \label{1}
\end{equation}
 where $V(q)=p^q=\cos ^{2q}(\Omega \tau /2)$ is the conditional
probability of the ion remaining in its original eigenstate under $q$
attempts of coherent excitation or deexcitation. In contrast, completely
preserved correlations $-$ equivalent to the absence of ``state reduction''
or ``collapse'' $-$ would require $V_{coh}(q)=\cos ^2(q\Omega \tau /2)$.

Relaxation modifies the probabilities $p_i$. An analytic solution of Bloch's
equations on resonance [39] yields, for $\theta \gg \pi $, when dispersive
interaction is negligible, 
\begin{equation}
p_i=1-f_iB_i(1-e^{-(a+b)}\cos \theta )\text{ ,}  \label{2}
\end{equation}
 where $B_0=\frac{\Omega ^2/2}{\Omega ^2+\Gamma \gamma }$, $%
B_1=1-B_0$ , $2a=\gamma _{ph}\tau +(\Gamma /2)\tau $ , $2b=\Gamma \tau $, 
\mbox{$\theta ^2=(\Omega \tau )^2-(a-b)^2$} , $\Gamma $ is the decay rate of the
inversion, and \mbox{$\gamma _{ph}=(2a-b)/\tau $} is the rate of phase diffusion of
the drive light [40]. The factor $f_i$ takes into account the Zeeman
splitting in the ground state $(f_0=1/2)$, and preparation in the metastable
state with possibly mixed orientation $(f_1<1)$. From the {\it recorded}
spectrum, corresponding to the calculated one shown in Fig. 3, the nutation
phase \mbox{$\theta =2\pi n+\theta ^{\prime }$} was derived [25], where n was found
about 640, and $\theta ^{\prime }$ close to $\pi $ for peak values in the
spectrum, and close to zero in the dips, corresponding to maximum and
minimum probability of excitation. From the trajectories of results, the
numbers of sequences have been extracted that are made up of $q$ consecutive
equal results, $U(q)$. These quantities, normalized by $U(1)$, have been
compared with the joint probability $V(q)$, (Fig. 4). One of the
trajectories is required for determining, from the ``on'' sequences, the
parameter of total relaxation, $a+b$. The ``on'' sequences in all the
remaining trajectories allow the measurement of the fractional phase of
nutation, $\theta ^{\prime }$. Note that this phase is available with
substantial precision (see Fig. 4). The ``off'' sequences yield $f_1$. In
such sequences of ``no count'' observations, the factor $f_1$ decreases from
unity $(\theta ^{\prime }\simeq 0)$ upon increasing deexcitation (growing $%
\theta ^{\prime })$, since more and more cycles of spontaneous decay
followed by stimulated reexcitation contribute to the ``off'' results.

The agreement of $V(q)$ and the normalized $U(q)$ reveals the driven
evolution of the ion being set back during the action of each of the probe
pulses. This behaviour may or may not be interpreted by repeated
``reduction'' to the initial energy eigenstate, or ``collapse'' ([9]; see
Ref. 8 for a comprehensive discussion of the relevance of collapse to QZE).
However, there is no way to invoke dephasing, since the spin-equivalent
individual quantum object -- the driven quadrupole -- pertains to a pure ion
state and displays a well-defined phase except in the two eigenstates of
energy which are, however, generated only upon particular excitation, namely
by $\pi $ or $2\pi $ pulses of the drive light. Thus it seems that finally
the effect of measurement has been unequivocally unveiled from the disguise
as physical phase perturbation that it assumes when a quantum {\it ensemble}
is observed.

\section{Back-action on the quantum object}

An important issue concerning potential phase perturbation requires
attention: the amount of back action of the light fields upon the individual
quantum system in the course of the above procedure of measurement. A
sufficient -- although not a necessary -- condition for the exclusion of
large enough back action is the measurements being of QND type. An
observation qualifying for this category needs a quantum object whose state
has been entangled with a quantum probe that is subjected to the physical
detection [21]. The result of this detection permits one to infer the state
of the quantum object thanks to its correlation with the state of the
quantum probe. A sufficient condition for QND is 
\begin{equation}
{\bf U}^{+}\text{ }{\bf x}\text{ }{\bf U\ -}\text{ }{\bf x}\text{ \ }{\bf =}%
\text{ \ }0\text{ ,}  \label{3}
\end{equation}
where ${\bf x}${\bf \ }is the operator of the relevant observable,
and ${\bf U}$ is the operator of the joint time evolution of probe and
object. This condition demands that both quantum object and quantum probe
return to their respective states after the measurement. In the above scheme
of observation, the ionic quadrupole induced by the drive is the quantum
object, while the dipole on the resonance line is the quantum probe. Let us
scrutinize the possible outcomes of the outlined procedure of measurement.

There are two kinds of results that are characterized by probe-light
scattering ``on'' or ``off''. ``Off'' detections are unrelated with any
light scattering. Moreover, they do not cause any physical recoil at all on
the quantum system since the probe-line dipole and the concomitant resonance
scattering remain unexcited and establish a ``negative result''. Note that
this {\it absence} of light scattering allows one, thanks to the
entanglement with the quantum object, to infer upon the state of the quantum
object, i.e. the ion resting in the dark $D_{5/2}$ level.

``On'' results indicate interaction of ion and probe light, by the latter
inducing the oscillating dipole of the former on the probe resonance that
gives rise to the scattering. Since both quantum object and probe return to
their initial states after each cycle of measurement, the QND condition
seems to hold. But probe light scattering also gives rise to stochastic
momentum transfer to the ion. However, laser cooling establishes a
stationary vibrational state; and the temporal distribution of the ion in
phase space remains invariant, in agreement with the QND condition.

Now we should examine whether or not ``direct interaction'' [8], of the
probe light with the quantum object, may effect {\it physical} intervention
in the latter to be expressed as collapse, state reduction, or their
equivalent. Such an interaction might take place via either 1. electronic
excitation, or 2. light recoil:

1. According to the quantum system's interaction with the drive light, the
superposition state 
\mbox{$\alpha $ $\langle $ $n,1\mid $ + $\sqrt{1-\alpha
^2}\langle n+1,0\mid $} of the quantum object, with 
\mbox{$\alpha =\alpha
(\Omega \tau),$} is prepared and said to turn, by help of the probe light,
into $\langle $ \mbox{$n+1,0\mid ,$} where $n$ is close to the mean photon
number of the coherent drive light. After the drive light has been switched
off, and the probe light on, \mbox{$\langle 0\mid $} correlates with probe
light scattering ``on'', and \mbox{$\langle $ $1\mid$} with ``off'', such
that rather \mbox{$\langle $ ``on'', $n+1,0\mid $} results. But how does the
probe light manage to ``reduce'' the quantum system into state 
\mbox{$\langle $
``on''$,0\mid $} ?

For an answer, we should distinguish (\textit{i}) what we \textit{infer}
from the quantum-mechanical model designed for the description of an
ensemble: the preparation in the superposition state -- and (\textit{ii})
what we \textit{measure }and \textit{know} from the result of the
measurement, and to what we attribute ''reality'': the system found again
and again in the \mbox{$\mid 0$ $\rangle
$} state. This distinction is indispensable when dealing with single
measurements on an individual system. If we ignored this distinction and
attributed\textit{\ }reality to the predictions of quantum mechanics, when
applied to individual measurement on an individual quantum system, for the
time intervals between measurements, e.g., to the expectation value of
energy, then both the excitation of the dressed state, and the presumed
subsequent jump from the inferred left-over superposition state to %
\mbox{$\mid $``on''$,0$ $\rangle $} were identifiable \textit{physical
processes. }They were going on in the quantum object, induced by interaction
with drive and probe, respectively. This two-step scenario is impossible to
take place in an individual atom, for sake of the atom's quantized internal
energy. Moreover, the jump had to be represented by a Hamiltonian that
enters the deterministic Schr\"{o}dinger equation, although it is well
established that the nonlinear and stochastic collapse of the wave function
cannot be described by a linear, deterministic Hamiltonian interaction of
system and probe. In fact, any measurement following coherent excitation
results, with respective probability, in one of the \textit{eigen}values of
states 0 or 1. In a sequence of ''on'' observations, we know \textit{a
posteriori }that \textit{all} the results have been ''state 0'':
Consequently, there is no motive for attributing other states to the quantum
object (the driven resonance), between two measurements, and eq. 3 holds.
Thus, the quantum object remains free of reaction, and it is not reasonable
to invoke, for sequences of equal results, ''direct interaction'' with the
probe.

On the other hand, phase and amplitude noise that is imposed, by the
probe-light pulses, upon the ion's ground-state wave function makes the
quadrupole on the driven transition (the quantum object) {\it decohere},
although this effect cannot give rise to the above transitions.

In fact, we know from the series of QND measurements that the system keeps
being found in 
\mbox{$\mid $``on'',$ 0 \rangle $}, and {\it assuming } the system
to have been elsewhere between measurements is not substantiated, although
the {\it a priori} probability for the experimental finding has been less
than unity, according to preparation by the drive light. A measurement of
the state aquired by this preparation in a base of which that state is an
eigenstate would verify the result of preparation. If such a measurement
were sandwiched inbetween the drive and probe pulses, it would be left, by
the subsequent probing (measurement of energy state with result 
\mbox{$\mid 0\rangle $),} an {\it incompatible} measurement of a non-commuting variable.
The results of such two incompatible measurements cannot {\it simultaneously}
claim reality, a quality that depends on the selected base being adapted to
the observable measured. Consequently, there is no {\it physical} collapse
that would require ``direct interaction'', but rather reduction of
possibilities, i.e. enhancement of knowledge, by the measurement.

2. Scattering of the probe light may exert recoil to the quantum system
However, if the vibrational frequency $\nu _{\text{v}}$ of the ion in the
trapping potential exceeded the widths of {\it both } lines, the macroscopic
trap would absorb the recoil exchanged in the course of any radiative
interaction of ion and field, since this interaction extends over many
vibrational periods [41]. In the actual experiment, this ``strong trapping''
holds with the drive light, although it does not hold with the probe. At any
rate, the ion remains laser-cooled, while probed, and characterized by a 
{\it narrow} stationary distribution of its momentum. Detection of
individual acts of recoil could, in principle, replace the detection of the
scattered probe light, and so far this detection would amount to another way
of measuring the state ``on''. As well as the recording of scattered light,
this detection of momentum transfer cannot {\it physically} do anything like
changing the quantum object's superposition state into $\langle $ ``on''$%
,0\mid $. However, it will cause decoherence as a consequence of Doppler
phase modulation of the quantum object's wave function. Since the ion is
laser-cooled to an extent that leaves its vibrational excursion far smaller
than half the wavelength of the light (it remains in the ``Lamb-Dicke
regime''), the extent of this decoherence is rather limited. The phase
perturbation imposed on its wave function by the momentum transferred from
the probe light is restricted to just a negligible fraction of $\pi $ .
Consequently, this small phase perturbation of the quantum object does not
suffice for mimicking the effect of information-enhancing measurements, as a
consequence of the non-local correlation of the quantum system with the
result of counting, on the macroscopic level, the photo-electrons released
by the scattered probe light. By the same token, substantial decoherence via
phase-fluctuating signal light cannot be tolerated [25].

In summary, the actual conditions of the recent experiment, for both ``on''
and ``off'' results, in fact exclude physical intervention with the quantum
object -- the ion's quadrupole moment on the signal line -- as the origin of
the observed statistics of results.

\section{Exponential decay vs. coherent evolution}

Some objections have centered at questioning the relevance of a test of the 
{\it coherent} evolution of a quantum system as opposed to a test of the
spontaneous decay of such a system [24]. Indeed, the initial concept of the
retarded evolution identified with the quantum Zeno effect was concerned
with spontaneous decay of a quantum system [3-5]. As discussed above, the
argument for retardation to show up rests on the presumed existence of a
short initial time regime when the occupation density of the decaying state
decreases as the square of time, and specifically as a Gaussian. Such a
regime has been shown to emerge from an unstable quantum system under very
general conditions [3, 4]: The non-decay probability for a quantum system
decaying into a stable final state cannot, for an initial short time regime,
follow an exponential which is the signature of classical evolution [30].
Thus, the strange and possibly paradoxical aspects of QZE are constituents
of {\it any} physically unperturbed quantum evolution subjected to
intervention by observations, in particular of a quantum evolution where the
existence of that quadratic time dependence is obvious, as the coherent
evolution. Nonetheless, testing spontaneous decay for that quadratic regime
of evolution -- and the concomitant QZE -- could provide most valuable
insight into the emergence of classical behaviour in a compound of quantum
systems. Recent proposals for testing decay processes that even predict
accelerated decay [33, 34] are based on interaction of the quantum system
with a radiative reservoir, modelled in the approximation of Fermi's golden
rule. These approaches would correspond to a definition of QZE that does not
discriminate between physical back action and non-local correlation [8].
However, a test of spontaneous decay for initial deviation from exponential
behaviour would possibly allow one to check quantum models that differ with
respect to the reservoir, e.g. Breit-Wigner theory [42] and models like the
one detailed in Ref. 7. Desirable as such a test on a spontaneously decaying
quantum system seems -- it is evident that the uncovering of the effects of
such a transient regime requires either experimentation on an extremely
short time scale, and/or with excessively high sensitivity. Therefore, it is
not surprising that attempts have been surfaced in order to first manipulate
the exponential decay of the quantum system such as to make it display a
quadratic regime of decay that extends over a time interval more accessible
by current experimental technique [24]. In order to follow this strategy, a
supplementary coherent excitation is proposed to be applied to the quantum
system, and a nonlinear contribution to the interaction may be detected that
mixes the amplitudes of the coherent and dissipative parts. The evolution of
this nonlinear contribution is supposed to display, on an accessible
intermediate time scale, a t$^2$ evolution, or even some of the oscillatory
variation that goes along with the coherent interaction but is, at later
times, overwhelmed by the incoherent decay. However, this t$^2$ regime is
the signature of just the admixed coherent contribution to the interaction,
and it is by no means evident how to $\inf $er, from its observation, the
existence of a much shorter t$^2$ regime in the {\it incoherent}
interaction, let alone the physical origin of this transient dynamics. Thus,
the strategy of Ref. 24 is of questionable value for the decision whether or
not natural decay is accompanied by the deviation from exponential decay
being the condition of QZE. For the substantial relevance the answer to this
open question has upon the problem of the boundary between quantum
micro-system and classical macro-system, we suggest a possible pathway out
of the dilemma posed by the enormous sensitivity of detection required, as
it seems, for a meaningful test of the dynamics of a ``natural'' exponential
decay.

\section{A sensitive test of exponential decay being modified by
measurements}

The decay time related to a resonance as considered in the above experiment
[25] is $\Gamma ^{-1}=\tau /2b$, and $b$ is the quantity to be measured with
utmost sensitivity in order to reveal any variation of $\Gamma $ under more
frequent probing. A straight-forward approach would consist of increasing
the repetition rate of the drive and probe cycles, and check the
corresponding trajectories of measurements for a {\it variation }of $b$
derived from precise determinations of the fractional phase of nutation, $%
\theta ^{\prime }$. Although this approach seems feasible in principle, it
is yet unpractical: In fact, one natural mode of relaxation, characterized
by $a+b,$ is easily accessible from fitting $U(q)/U(1),$ the other one, $a-b$%
, is not, since it must be derived from a small difference of the large
quantities $\theta ^2$ and $(\Omega \tau )^2$. A better approach would rely
on the comparison of {\it correlated} fractional phases $\theta ^{\prime },$
as in interferometry, such that their relative variation is directly traced
back to a variation of the relaxation constant $b$. This strategy is
outlined in what follows.

The accumulation of trajectories of measurements may be modified such that
the probe light is alternated only after two, or three, or $n$ pulses of the
drive. The corresponding phases of nutation are 
\begin{equation}
\theta _n=\sqrt{n^2(\Omega \tau )^2-(a-b_n)^2}=2\pi ns+\theta _n^{\prime }%
\text{ \ ,}  \label{4}
\end{equation}
 where $s$ is the number of integer nutational rotations induced by
one driving pulse. The anticipated retardation of the quantum evolution
would make $b_n$ decrease upon decreasing $n$. With $\Omega \tau >\pi $, the
square root is expanded in orders of $\frac{a-b}{n\Omega \tau }$ . The
fractional phases $\theta _n^{\prime }$ may be determined with less than $%
10^{-4}$ error, and one derives, with $b_m=b_n-\delta b$, 
\begin{equation}
\delta _{mn}\equiv \frac{\theta _m^{\prime }}m-\frac{\theta _n^{\prime }}%
n=\frac 1{\Omega \tau }\left( \frac{a-b_n}n\right) ^2\left[ 1-\left( \frac
nm\right) ^2\left( 1+\frac{\delta b}{a-b_n}\right) ^2\right] \text{ .} 
\label{5}
\end{equation}
 From the value $\delta _{m1}$ derived from $\theta _m^{\prime }$
and $\theta _n^{\prime }$ of trajectories at $n=1$ and $1\ll m$ $<\Gamma
^{_{-1}}/\tau ,$ one derives $a-b_1$, with $\Omega \tau $ taken from fitting
a spectrum of the excitation, or from calibration of the light flux .

A measured value of $\delta _{21}$, at $n=1$ and $m=2,$ allows us to
determine the presumed variation $\delta b$, by effectively doubling the
driving period. This may be illustrated by a numerical example: With the
realistic values $a\simeq 0.4,$ $b_1\simeq 0.2,$ $\Omega \tau
\simeq 2\pi ,$ we have $\delta _{m1}\simeq 6\times 10^{-3}\pm
2\times 10^{-4},$ and $\delta _{21}=\delta _{m1}\left( 1-\frac 14\left(
1+5\delta b\right) ^2\right) \simeq \delta _{m1}\left( \frac 34-\frac{10}%
4\delta b\right) .$ Thus, the maximum error of $\delta _{21}/\delta _{m1}$
is $4\times 10^{-4}$, and the error of $\delta b$ would amount to $1.6\times
10^{-4}$. This level determines the limit of sensitivity for the observation
of modified {\it decay }with the experiment described above [25].

The outlined strategy represents a kind of ``heterodyne'' measurement of the
effective fractional phase of nutation. As a ``local oscillator'', the
corresponding phase of the coherent evolution is made use of.

For an experimental demonstration of any observation-induced variation of
the rate $b$ of energy relaxation, the excited state of the driven resonance
must live long enough to exceed a rather long sequence of, say, ten of the
standard drive/probe periods, and short enough to allow the detection of the
reduced decay parameter $b$. The $S_{1/2}-D_{5/2}$ transition of $%
^{172}Yb^{+}$ with its lifetime of the $D_{5/2}$ level on the order of $5$ $%
ms$ [43] may in fact offer a good compromise. An experiment along these
lines would yield at least an upper boundary for the measurement-induced
retardation (or even acceleration) of the exponential decay, although this effect 
may be still several orders of magnitude inferior. 

\section{Conclusions}

The conditions for demonstrating the inhibition or retardation of the
quantum evolution of an atomic system by reiterated measurements have been
reviewed in the light of various arguments that had been raised with respect
to an attempted demonstration some ten years ago [10], and on account of a
recent experiment on an individual atom [25].

Questions about the general applicability of von Neumann's principle of
state reduction already have been clarified when a single-particle model
calculation proved this principle to provide an excellent approximation to
the ideal evolution of the system [9]. Moreover, the existence or
nonexistence of state reduction and collapse have been shown irrelevant to
the problem of the measurement-affected evolution [8].

In order to identify the retarding effect of observation by a quantitative
evaluation, it seems indispensible to {\it register} the results of all the
interrogations. Otherwise, correlated up-down transitions in pairs of atoms,
or back-and-forth transitions in a single atom, would falsify the crucial
probability for no transition that is derived from the observations at least
in a real experiment [17].

The most serious objections have been aimed on the very nature of the
impeding effect. It has been argued that, for a signature, the measurements
should be nonlocal and of the negative-result type [8]. Although we
acknowledge this requirement as a sufficient condition, there is good reason
to accept, as ``good'' measurements, QND measurements with positive results,
and non-QND measurements as long as the back action {\it demonstrably}
cannot cause the inhibition of evolution.

The impeding effect of {\it measurement }has been, in the meantime, proven
indistinguishable from dephasing the system's wave function, {\it except}
for an individual particle [18-23]. An experiment on such a system, a single
trapped and laser-cooled ion, including the read-out of the results of {\it %
all} interactions with the probe light, has been reported recently [25]. The
results of this experiment show that the observed retardation of the
system's evolution is unequivocally traced to the repeated interrogation of
the quantum system's internal state of energy, and {\it not} to dephasing of
the system's wave function.

The controversy also dwelled upon the experimental test addressing {\it %
coherent} evolution as opposed to exponential decay [24]. In fact, unstable%
{\it \ } quantum systems are supposed to include an initial very short
regime of quadratic time evolution that is prerequisite also of the ``true''
Zeno variant of quantum-evolutive retardation. If a system lacked this
regime, this finding would dramatically highlight the discrepancy between a
quantum micro-system, and such a classical macro-system emerging from the
action of a reservoir. As for a proof, the admixture of a coherent
contribution to the system's evolution as previously suggested [24] is of
questionable evidential value. On the other hand, the anticipated time delay
of an exponential decay may be determined, with substantial precision, by a
modification of the reported drive-probe experiment. At least a meaningful
upper bound for that elusive retardation might be hoped for.

The evolution of a quantum system is always predicted, with the help of the
deterministic Schr\"{o}dinger equation, on the base of previous measurements
or assumptions, that determine the initial conditions. Even the {\it %
availability} of additional information on the system requires partial
updating, on the ground of a potential measurement having happened. The
retardation of the quantum evolution is the signature of this continuated
process of updating. The predictable quantum evolution conditions the
stochastic ``factual'' evolution of the system, at least in the
probabilistic interpretation of quantum mechanics. If this is so, the
Quantum Zeno Effect emerges as a consequence of all, that we can know about
a quantum system's evolution, remaining incomplete.

\end{multicols}{2}

\newpage

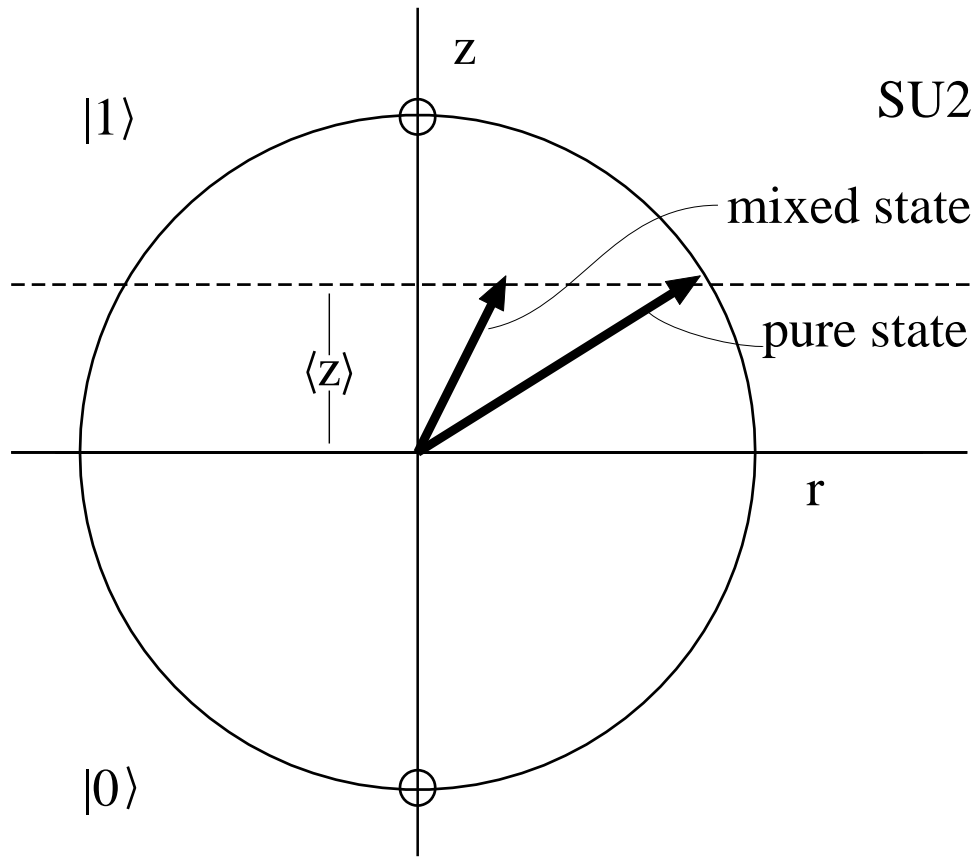
\begin{figure}
\caption{The configuration space of an ensemble of two-level atoms
fills surface {\it and} interior of the Bloch sphere. A measurement yields
an expectation value $\langle ${\bf z}$\rangle $ with $0\leq \langle ${\bf z}%
$\rangle \leq 1$. The configuration space of an individual such atom is
restricted to the surface. A measurement of internal energy yields the
eigenvalue 0 or 1, and a series of measurements, after equal preparation,
reveals the micro-state that is a pure one.} 
\label{Fig1}
\end{figure}

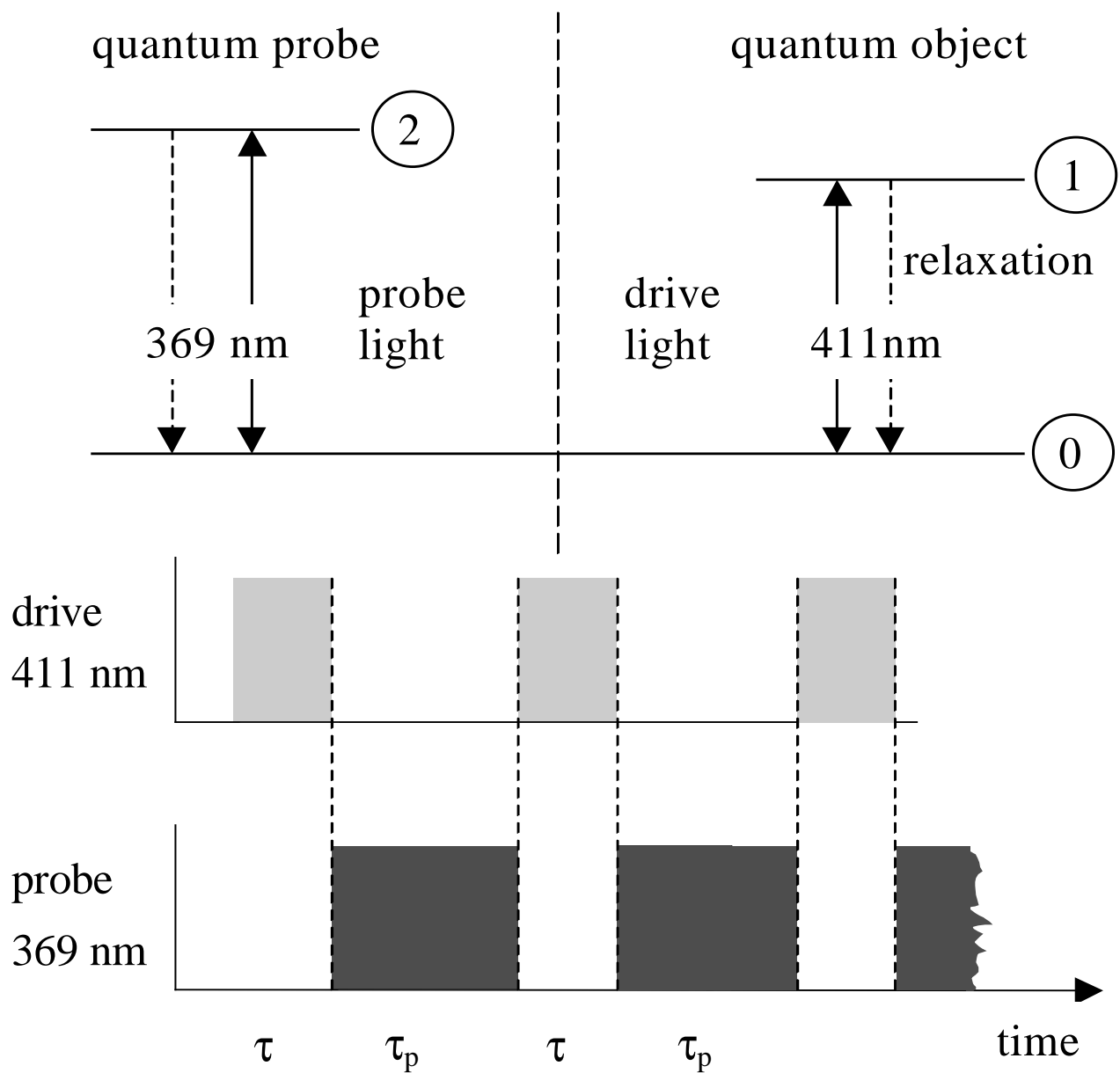
\begin{figure}
\caption{Level scheme of an ion ($Yb^{+}$; level 0: $S_{1/2},$1: $%
D_{5/2,}$ 2: $P_{1/2}$) interacting with monochromatic drive light resonant
with $E2$ transition, and with probe light that excites light scattering on
a resonance transition.} 
\label{Fig2}
\end{figure}

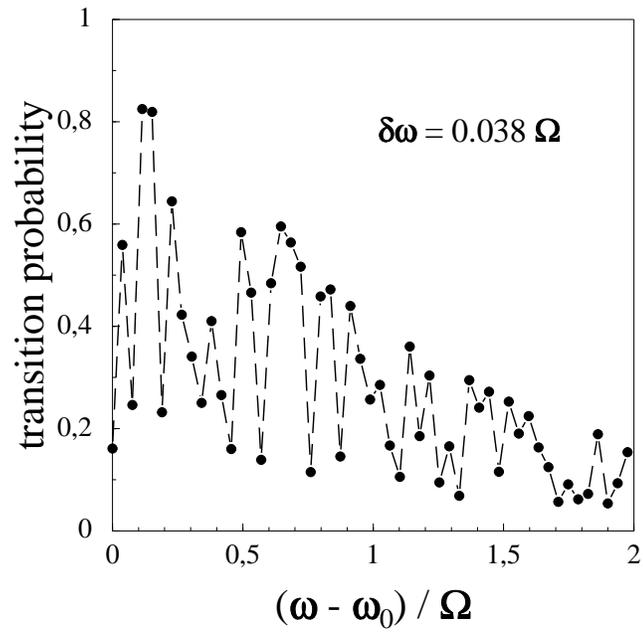
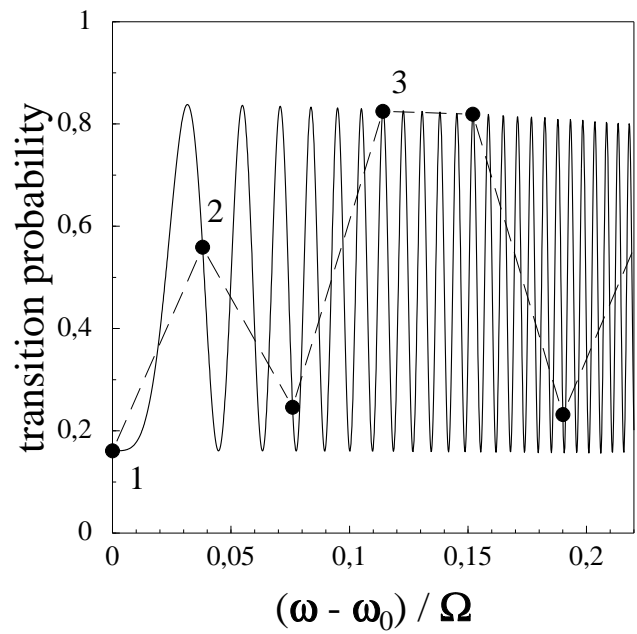
\begin{figure}
\caption{Excitation probability $p_{01}=1-p_0$ on coherently driven
transition 0-1, equivalent to eq. 2, but simulated from numerical evaluaton
of Bloch equations. Top: The spectrum demonstrates the generation of the
observed data of Fig. 3, Ref. 25 (Rabi frequency $\Omega =500$kHz$\times
2\pi )$. Bottom: Expansion of the scale of detuning reveals Rabi nutation
and the stroboscopic sampling with step size $\delta \omega .$ The marks 1,
2, and 3 indicate transition probabilities that correspond to trajectories
of data evaluated in Figure 4.} 
\label{Fig3}
\end{figure}

\begin{figure}
\caption{ Probabilities $U(q)/U(1)$ of uninterrupted sequences of $q$
``on'' results (white dots) and ``off'' results (black dots). The lines show
the distributions of probabilities $V(q-1)$ for the ion's evolution on its
drive transition, according to Eqs. 2 and 3. $\theta ^{\prime }$ and $f_1$
from fit; values $f_1<1$ indicate redistribution, over sublevels, by cycles
of spontaneous decay and reexcitation. From Ref. 25.} 
\label{Fig4}
\end{figure}

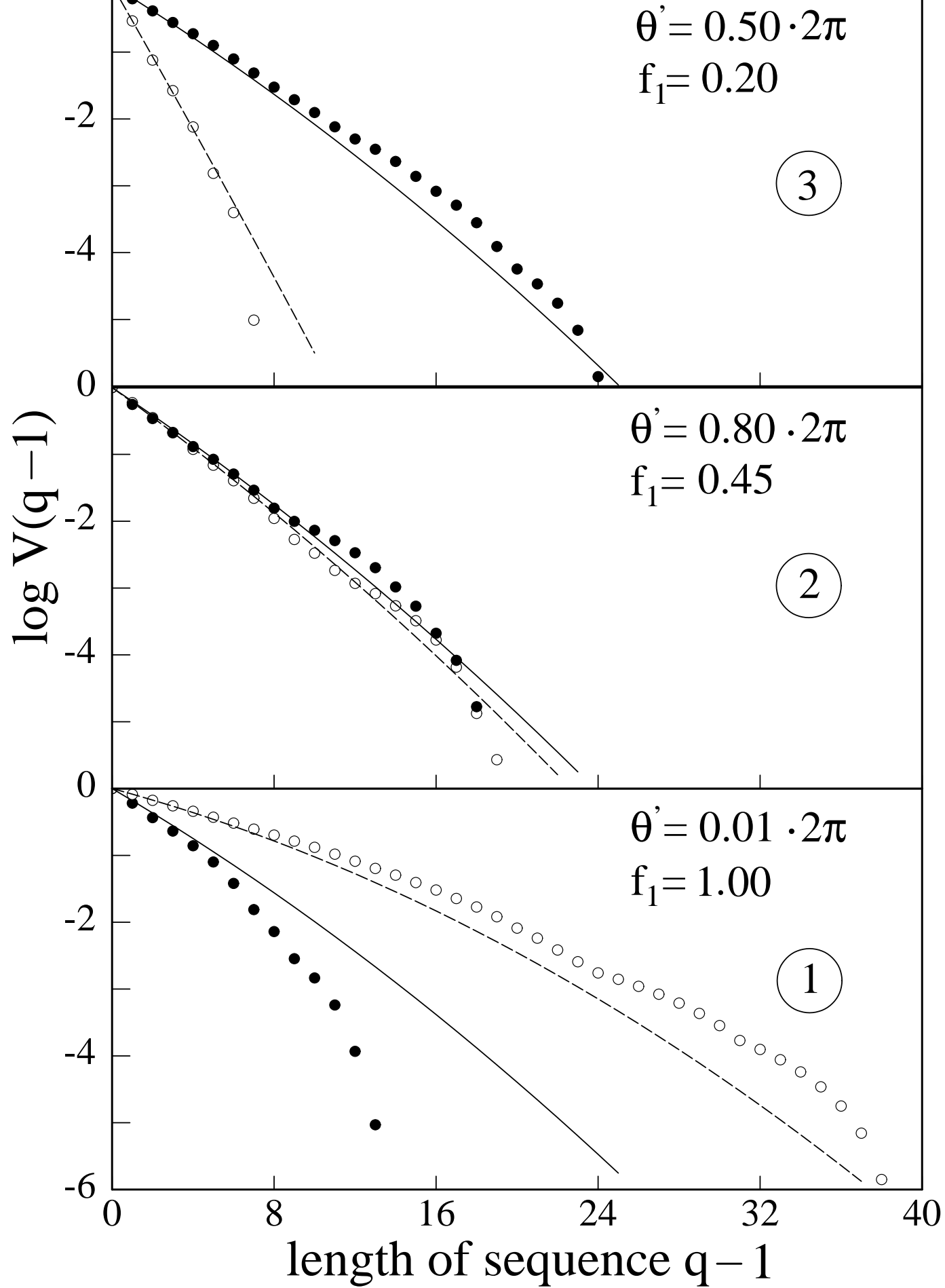
\end{document}